\documentclass[aps,prl,reprint,superscriptaddress]{revtex4-1}
\usepackage{lipsum}
\usepackage{graphicx}
\usepackage{amsmath,amssymb}
\usepackage{bm}
\usepackage{dcolumn}
\usepackage{theorem}
\usepackage{wrapfig}
\usepackage{mathtools}
\usepackage{xcolor}
\usepackage{txfonts}

\theoremstyle{break}

\newcounter{mycounter}
\newcommand{\state}{Statement}
\newtheorem{theorem}{\state}

\theoremstyle{break}

\begin{document}

\title{Ultrarelativistic limit of a Rarefied  Gas with Internal Structure}

\author{Sebastiano Pennisi$^1$, Tommaso Ruggeri$^2$}
\email[]{spennisi@unica.it, tommaso.ruggeri@unibo.it}

\affiliation{$^1$ Department of Mathematics and Informatics, University of Cagliari, Cagliari (Italy)\\
	$^2$Department of Mathematics and Alma Mater Research Center on Applied Mathematics AM$\,^2$, University of Bologna, Bologna (Italy)}

\date{\today}
\begin{abstract}
The aim of this letter is to check the ultra-relativistic limit of a recent model  proposed by Pennisi and Ruggeri [Ann. Phys.  377,  414  (2017)]
for a  relativistic gas with internal structure. Considering an  Eulerian  fluid   we prove that there exists a critical value of the degree of freedom such that for smaller values of this quantity the ultra relativistic  limit of  the energy of  a gas with structure  is the same as the Synge energy for monatomic gases, while for larger  degrees of freedom the  energy  increases with the degree of freedom itself. 
The limiting equations are accompanied with the explicit expression of the  characteristic velocities of the hyperbolic system.  
\end{abstract}



\maketitle



\section{Introduction}
In a recent paper Pennisi \& Ruggeri have presented a  casual relativistic theory for a non-equilibrium rarefied gas with internal structure \cite{Pennisi_Ruggeri}. The mesoscopic justification of the theory is the relativistic kinetic theory with a distribution function that, as in the classical case, depends  on an additional continuous variable 
representing the energy of the internal modes. This permits the theory to take into account the energy exchange between translational modes and internal modes of a molecule in binary collisions. The theory includes the ET theory of relativistic  gases by Liu-M\"uller-Ruggeri  \cite{LMR,RET} as a singular limit and in the classical limit converges to the model of Extended Thermodynamics for a polyatomic gas \cite{Arima-2011,Ruggeri_Sugiyama}. Therefore we have called the theory \emph{Relativistic  Extended Thermodynamics of rarefied  polyatomic gas}. In reality, as well written in the recent book by Rezzolla and Zanotti 	\cite{Rezzolla} "\emph{relativistic fluid is constructed around the concept of a fluid as a system whose large-scale properties can be described effectively without having to worry about the features that the constituent elements have at much smaller length-scales}".  
 Thus we expect that our theory is a  more refined   model, able to account also for the internal motion. The theory  is motivated at the kinetic level by a  modified J\"uttner   equation in which the distribution function  depends also on an internal energy that takes into account the internal motion (rotation and vibration)  \cite{Pennisi_Ruggeri}.
 In this letter we want to analyze in the simple case of non dissipative gas what happens in the ultra relativistic limit i.e  when the ratio
\begin{equation*}
 \gamma = \frac{mc^2}{k_B T}
\end{equation*}
 is very small  ($m$ is the particle mass, $c$ the light velocity, $k_B$ the Boltzmann constant and $T$ the absolute temperature) 
 and to compare our theory with the one for a relativistic  Eulerian fluid equipped  with Synge energy.
 In this limit   the bodies are so extremely hot that the mean kinetic energy of particles surpasses  the rest energy    or the mass is extremely small.
This condition is particularly interesting in some limiting  cases of nuclear physics or in some extreme astrophysical problematics.  \cite{Synge,anile,Kremer,Rezzolla}.

\section{The Relativistic  non dissipative model of rarefied  gas with internal structure } 
Let  us  consider a non dissipative gas based on 
the conservation laws of particle  number 
and energy momentum
\begin{equation}\label{1}
\partial_\alpha V^\alpha =  0   \quad , \quad
\partial_\alpha T^{\alpha \beta} =  0,
\end{equation}
where $\partial_\alpha = \partial/\partial x^\alpha$ with $x^\alpha$ being the space-time coordinates $\alpha =0,1,2,3$.
\begin{equation} \label{vatab}
V^\alpha =  n  m U^\alpha,   \quad  \quad
T^{\alpha \beta} =   p  h^{\alpha \beta}  + \frac{e}{c^2} \, U^{\alpha } U^\beta,
\end{equation}
where   $U^\alpha$ is the four-velocity ($U^\alpha U_\alpha = c^2$), $n$ is the number density, $p$  is the pressure, 
$h^{\alpha \beta}$ is the projector tensor:
\begin{equation*}
h^{\alpha \beta}= -g^{\alpha \beta} +
\frac{1}{c^2}  U^{\alpha } U^{\beta},
\end{equation*}
$g^{\alpha \beta}= diag(1 \, , \, -1 \, , \, -1 \,
, \, -1)$ being the metric tensor,   $e$ is the energy.
In this case  $T$,
$n U^\alpha$, are assumed to be  independent variables. 
For a fluid without internal structure, as it is well known,  the equations \eqref{1} can be obtained as the first $5$ moments of the 
  Boltzmann-Chernikov equation
\begin{equation}\label{B-C}
p^\alpha \partial_\alpha f =  Q,
\end{equation}
($p^\alpha $ is the four-momentum with the property $p^\alpha p_\alpha=m^2 c^2$, and  $Q$ is the collisional term) when we suppose that the distribution  function $f(x^\alpha,p^\alpha)$ is coincident with the equilibrium  J\"uttner distribution function:
\begin{equation*}\label{fJ}
f_J= \frac{n \gamma}{ K_2(\gamma)} \frac{1}{4 \pi m^3
	c^3} e^{- \frac{\gamma}{mc^2}   U_\beta
	p^\beta, }
\end{equation*}
where $K_m(\gamma)$ denotes the Bessel function of second kind.
In correspondence with the J\"uttner distribution function  we have the following expression for pressure and energy:
\begin{align}
\begin{split}
&   p =  \frac{m nc^2}{\gamma} = \frac{k_B}{m}\rho T,  \quad
 e= \frac{n m c^2}{  K_2(\gamma)}  \left[ K_3 \left(\gamma  
\right) - \frac{1}{\gamma}  K_2
\left(\gamma   \right) \right]. \label{caloric}
\end{split}
\end{align}
The expression of energy in \eqref{caloric} is called the Synge  energy  \cite{Synge,anile,Rezzolla}.

In \cite{Pennisi_Ruggeri} starting from the classical idea for polyatomic gas introduced  in \cite{Borgnakke-1975,Bourgat-1994, Pavic-2013,AnnalsMany,AnnalsLimit, Ruggeri_Sugiyama} we proposed a generalized Boltzmann-Chernikov equation \eqref{B-C} for  the extended distribution function $f \equiv f(x^\alpha,p^\beta,\mathcal{I})$.  
 By analogy with the classical case we consider the following moments:
\begin{align}\label{14n}
\begin{split}
&   V^\alpha = m c \int_{\Re^3} \int_0^{+\infty} f p^\alpha
\phi(\mathcal{I}) \, d \vec{P} \, d \, \mathcal{I}, \\
& T^{\alpha \beta} =
\frac{1}{mc} \int_{\Re^3} \int_0^{+\infty} f \left( mc^2 + \mathcal{I} \right) p^\alpha
p^\beta \, \phi(\mathcal{I}) \, d \vec{P} \, d \,
\mathcal{I},
\end{split}
\end{align}
where
\begin{equation*} 
d \vec{P} =  \frac{dp^1 \, dp^2 \,
	dp^3}{p^0} .  
\end{equation*}
The form of these equations is dictated by analogy with the classical case in which
it was necessary to introduce a distribution function with an extra variable
taking into account the internal degrees of freedom of a molecule. 

The meaning of $\eqref{14n}_{2}$ is that the energy and the momentum in relativity are
components of the same tensor and we expect that, besides the energy at rest, there is a contribution 
due to the degrees of freedom of the  gas because of  the internal structure, as in the case of a classical polyatomic gas.
$\phi(\mathcal{I})$ is the state density of the internal mode, that is, $\phi(\mathcal{I}) \, d  \mathcal{I}$ represents the number of the internal states of a molecule having the internal energy between $\mathcal{I}$ and $\mathcal{I}+d \mathcal{I}$.

In the classical limit, when $\gamma\rightarrow \infty$, the internal energy 
\begin{equation}\label{internalenergy}
\varepsilon = \frac{e}{m n} -c^2,
\end{equation}
converges to the one of a polyatomic gas:
\begin{equation}\label{DDDD}
\lim_{\gamma \rightarrow \infty} \varepsilon = \frac{D}{2}\frac{k_B}{m} T,
\end{equation}
provided that the measure 
\begin{equation}\label{phia}
\phi(\mathcal{I}) = \mathcal{I}^a,
\end{equation}
where the constant
\begin{equation}\label{aaa}
a= \frac{D-5}{2},  
\end{equation}
and $D= 3 + f^i$ is related to the degrees of freedom of a molecule given by the sum of the space dimension $3$ for the translational motion and the contribution from the internal degrees of freedom $f^i \geq 0$ related to the rotation and vibration. For monatomic gases $D=3$ and $a=-1$.
In the following we assume,  that the measure remain the same form as \eqref{phia} for any $\gamma$. 
 This choice is justified mathematically  in \cite{Pennisi_Ruggeri}.

In \cite{Pennisi_Ruggeri}  it was first  considered a non dissipative gas of Euler type
and it was required  that the generalized entropy
\begin{equation*}
\rho S =  - k_B \, c \,  U_\alpha \int_{\Re^3}
\int_0^{+\infty} f
\ln f p^\alpha \phi(\mathcal{I})  \, d \vec{P} \, d \, \mathcal{I}
\end{equation*}
has a maximum under the constraint that the $5$ moments \eqref{14n} are prescribed.  In this way in \cite{Pennisi_Ruggeri} the present authors founded the   equilibrium distribution function  that generalizes the J\"uttner one:
\begin{equation}\label{5.2n}
{f_E= \frac{n \gamma}{A(\gamma) K_2(\gamma)} \frac{1}{4 \pi m^3
c^3} e^{- \frac{\gamma}{mc^2} \left[ \left( 1 + \frac{\mathcal{I}}{m c^2} \right) U_\beta
p^\beta \right]}},
\end{equation}
with $A(\gamma)$ given by 
\begin{equation*}
A(\gamma)=  \frac{\gamma}{K_2(\gamma)} \int_0^{+\infty} \frac{K_2( \gamma*)}{\gamma*} \,   
\phi(\mathcal{I})  \, d \, \mathcal{I}, 
\end{equation*}
where 
\begin{equation*}\label{gammastar}
\gamma^*= \gamma + \frac{\mathcal{I}}{k_B T}.
\end{equation*}
The pressure and the energy for polyatomic gases, compatible with the distribution function \eqref{5.2n} are \cite{Pennisi_Ruggeri}:
\begin{align}\label{3n}
\begin{split}
&      {p =  \frac{n m c^2}{\gamma} = \frac{k_B}{m} \rho T}  \, , \\
&   {e= \frac{n m c^2}{A(\gamma) K_2(\gamma)} \int_0^{+\infty} \left[ K_3  (\gamma^*)   - \frac{1}{\gamma^*}  K_2
 (\gamma^*) \right]  \phi(\mathcal{I}) \, d \,\mathcal{I}}. 
\end{split}
\end{align}
We remark that the pressure has the same expression for a monatomic and for a polyatomic gas, while   \eqref{3n}$_2$ is the generalization of the Synge energy to the case of polyatomic gases.
When the measure $\phi(\mathcal{I})$ coincides with the delta of Dirac $\phi(\mathcal{I})= \delta(\mathcal{I})$ then \eqref{3n} converges to the relation for a monatomic gas \eqref{caloric}.
\begin{figure}
	\centering
		\vspace{-1cm}
	\includegraphics[width=.9\linewidth]{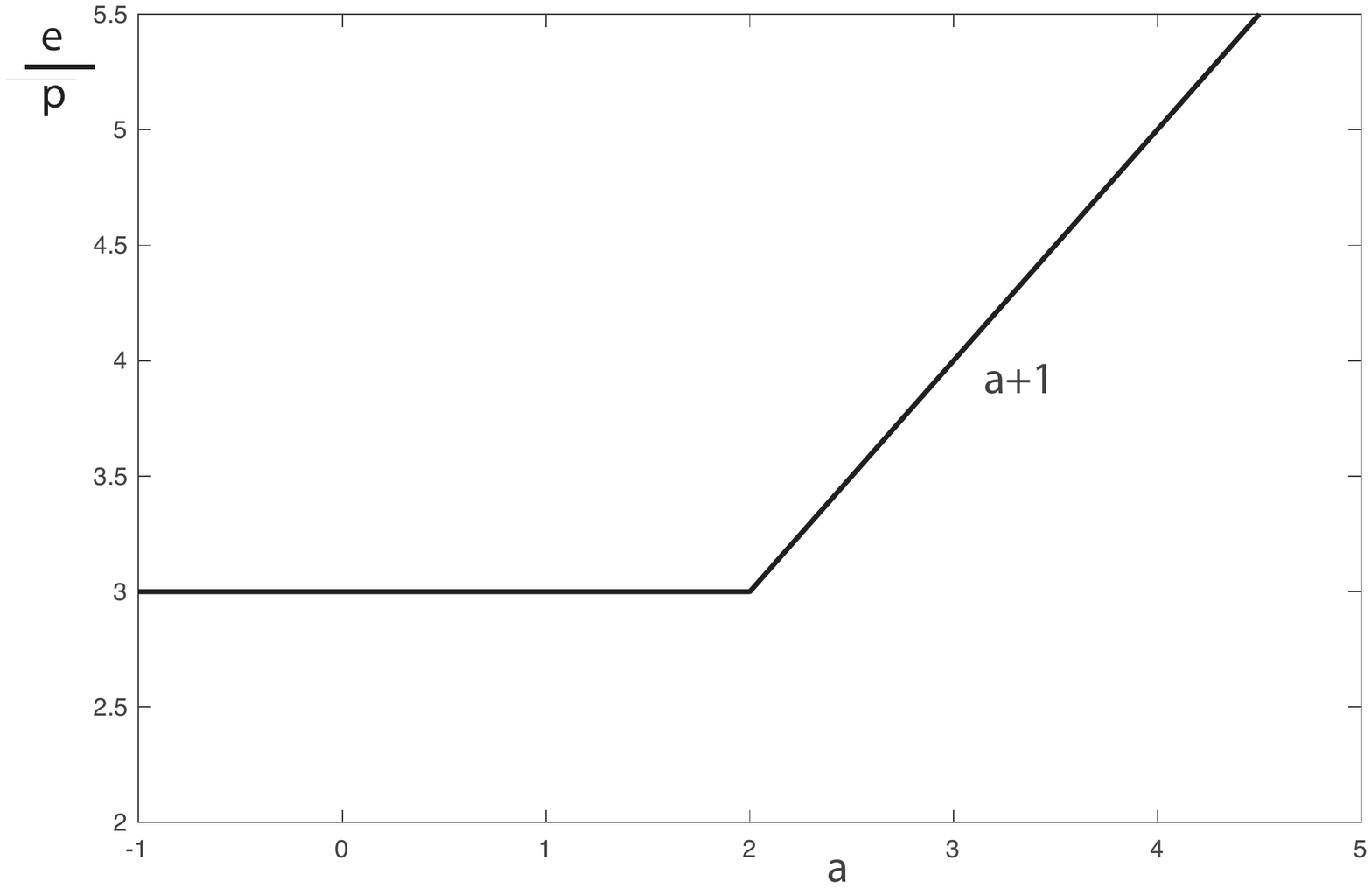}
	\vspace{-3cm}
	\caption{$e/p$ in the ultrarelativistic limit $ \gamma \rightarrow 0$ as a function of $a$.}
	\label{fig:energiapoly}
\end{figure}

\section{The ultrarelativistic limit for non dissipative polyatomic gas}
In the ultra-relativistic limit $\gamma \rightarrow 0$ the Synge energy for a monatomic gas \eqref{caloric}$_2$, converges,  as well-known, to 
\begin{equation*}
\lim_{\gamma\rightarrow 0} e_{\text{Singe}} = 3 n k_B T = 3p.
\end{equation*}
This result is obtained  by \eqref{caloric}$_2$ taking into account that  for $\gamma\ll 1$, we have 
\begin{equation*}
 K_3(\gamma) \sim \frac{8}{\gamma^3}, \quad 
 K_2(\gamma) \sim \frac{2}{\gamma^2}.
\end{equation*}
Now we want to analyze the ultrarelativistic limit for the energy of a polyatomic gas   \eqref{3n}$_2$, with $\phi(\mathcal{I})$ given by \eqref{phia}.
We prove the following:
	\begin{theorem}
		The energy of  gas with structure  \eqref{3n}$_2$ in the ultra-relativistic limit converge to
		\begin{align*}
	&	\lim_{\gamma\rightarrow 0} e = 3 n k_B T =3 p\quad \forall \, -1<a\leq 2, \quad \text{ and } \\
	&	\lim_{\gamma\rightarrow 0} e = (a+1) \, n k_B T =(a+1) \, p \quad \forall \, a > 2.
		\end{align*} 
	\end{theorem}
\begin{figure}
	\centering
	\includegraphics[width=0.9\linewidth]{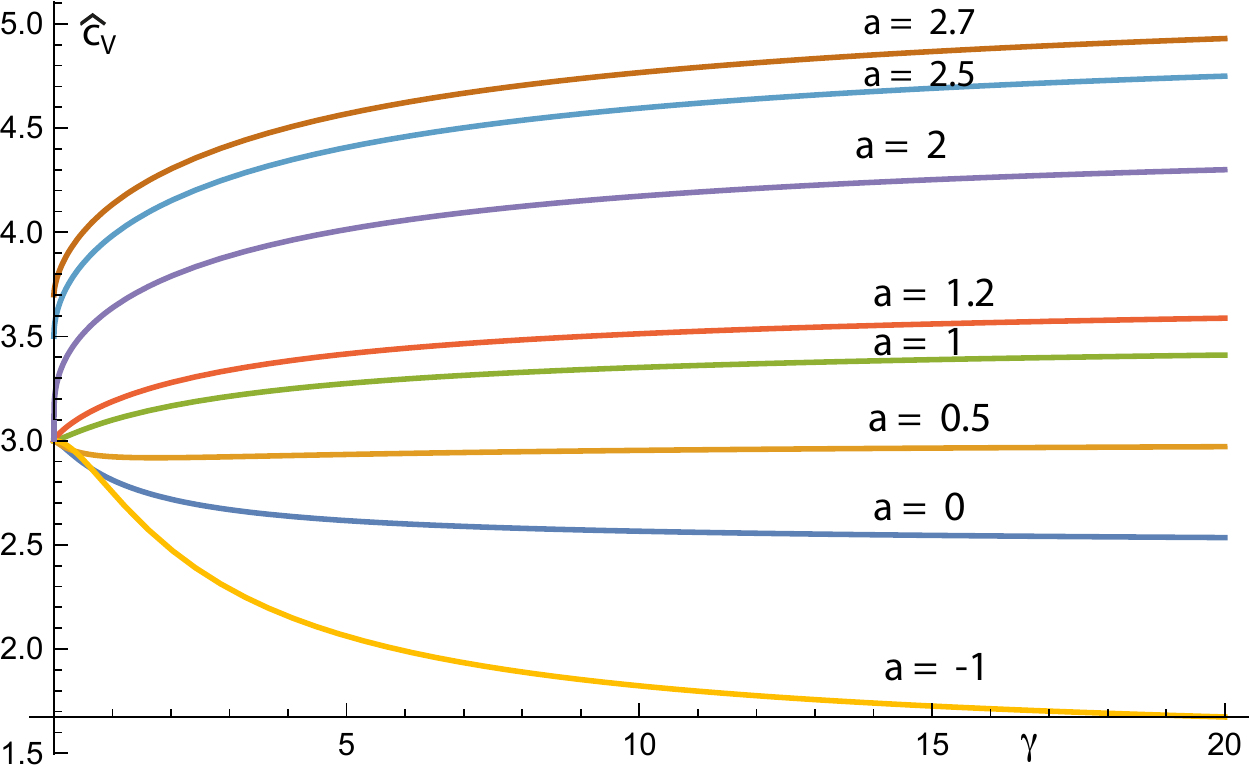}
	\caption{Specific heat $\hat{c}_V$ as function of $\gamma$ for different values of $a$.}
	\label{fig:specific}
\end{figure}
The proof can be done analytically but 
it requires some non trivial calculations  that we will present elsewhere. Here we refer to numerical results (see Fig.\ref{fig:energiapoly}) that are in perfect agreement with the statement. 
We notice from Fig.\ref{fig:energiapoly}, as we expect, that the limiting energy is the same as the limit of the Synge energy for a monatomic gas, not only for $a \rightarrow -1 $, but also for any $-1<a \leq 2$, i.e for moderate degrees of freedom. This fact seems to indicate that in the ultra-relativistic regime, due to the dissociation, any polyatomic gas becomes a sort of   monatomic one. On the other hand, the surprising result is that for $a>2$ this is not anymore true and the energy becomes more larger, increasing linearly with $a$. This result is difficult to interpret from the physical point of view because from one side we imagine that stable polyatomic gases do not exist for so large temperature. On the other hand, we cannot exclude that in some nuclear physics or astrophysics problems a stable  gas with internal structure  may exist also in the ultra-relativistic limit with very high temperatures or small mass if the degree of freedom are enough large!

If we differentiate eq. \eqref{internalenergy} with respect $T$ we obtain the specific heat $c_V$ as a function of $\gamma$. In Fig. \ref{fig:specific} we plot the dimensionless specific heat $\hat{c}_v= c_v m/k_B$ as function of $\gamma$. For large $\gamma$ corresponding to the classical regime it converges to the one of a polyatomic gas $\hat{c}_V=D/2 =a + 5/2$ (see \eqref{aaa},\eqref{DDDD}). While for $\gamma \rightarrow 0$, according to the previous statement it converges to $3$ for $-1 \leq a \leq 2$ and to $a+1$ when $a > 2$. Moreover, we  Fig. \ref{fig:specific} shows a strange behavior.  In fact, for  $-1 \leq a < 1/2$ the classical limit is smaller than the ultra-relativistic one and  
the specific heat is a decreasing function of $\gamma$, i.e., an increasing  function of the temperature. For $a = 1/2$ the two limits ultra and classical one are the same and therefore the function decreases, then has  a minimum and finally increases. After  $a> 1/2$ the classical limit is greater than the ultra-relativistic one and the specif heat 
 changes behavior and becomes an  increasing function of $\gamma$, i.e., a decreasing function of $T$.  We hope that there exist experimental data that can confirm this unusual behavior. We notice that the system is in any way hyperbolic and thermodynamically stable, because $c_V>0$ and the characteristic velocity are finite and smaller than the light velocity as we will see in the next section.

\section{Characteristic velocities in the ultra-relativistic limit}
Let $\lambda$ be the characteristic velocities of the differential system in light speed unity. We want to prove the
\begin{theorem}
	The characteristic velocities in the ultra-relativistic limit are:
\begin{align*}\label{case4}
\begin{split}
& \lambda =0 \quad \mbox{with multiplicity} \,\, 3 \\
& \lambda  = \pm { \frac{1}{\sqrt{r}}} \quad \mbox{each with multiplicity} \,\, 1  \, \\ 
& \quad \quad \quad \mbox{where} \quad r= \frac{e}{p}= \left\{ \begin{array}{l}
3 \quad \mbox{if} \quad a \leq 2 \\
a+1 \quad \mbox{if} \quad a >2.  
\end{array}
\right..
\end{split}
\end{align*}
\end{theorem}

\textbf{Proof}:
The balance equations \eqref{1}, \eqref{vatab} can be rewritten
\begin{align*}
\left\{ \begin{array}{l}
\partial_\alpha \left( n U^\alpha \right) =0 \, , \\
\\
\partial_\alpha \left[ n k_B T \, \left( h^{\alpha \beta} + \frac{r}{c^2} U^\alpha U^\beta \right) \right]=0  \, .
\end{array}
\right.
\end{align*}
The wave equations for the above system can be obtained as is well known with the chain rule:
\begin{equation*}
\partial_\alpha \rightarrow \left( \nu_\alpha - \frac{\lambda}{c} U_\alpha \right) \delta,
\end{equation*} 
where $\nu_\alpha$ is a four-vector restricted only by the conditions $\nu_\alpha \nu^\alpha = -1$ and $\nu_\alpha U^\alpha = 0$.
Therefore we have:
\begin{align}\label{wave}
\left\{ \begin{array}{l}
\left( \nu_\alpha - \frac{\lambda}{c} U_\alpha \right) \, \delta  \left( n U^\alpha \right) =0 \, , \\
\\
\left( \nu_\alpha - \frac{\lambda}{c} U_\alpha \right) \, \delta   \left[ n k_B T \, \left( h^{\alpha \beta} + \frac{r}{c^2} U^\alpha U^\beta \right) \right]=0 \, .
\end{array}
\right.
\end{align}
If $\lambda=0$, the eqs. \eqref{wave} become 
\begin{equation*}\label{wave1}
\nu_\alpha  \, \delta   U^\alpha  =0 \quad , \quad \delta  (nT)=0 \, . 
\end{equation*}
So there remains free one of the unknowns $\delta  n$, $\delta  T$ and the two components of $\delta   U^\alpha$ orthogonal to $U^\alpha$ and to $\nu^\alpha$. We conclude that $\lambda=0$ is an eigenvalue with multiplicity $3$. 
Taking into account that $r$ is constant in the ultrarelativistic limit for the result of Statement 1, 
if $\lambda \neq 0$ the eqs. \eqref{wave} become 
\begin{align}\label{wave2}
\begin{split}
& - \lambda c \, \delta   n + n \,  \nu_\alpha  \, \delta    U^\alpha  =0 \quad , \\
& -\nu^\beta \, \delta  (nk_BT) + nk_BT U^\beta \frac{r+1}{c^2} \nu_\alpha  \, \delta   U^\alpha - \\
&\frac{\lambda}{c} \left[ r U^\beta \, \delta   (nk_BT) + nk_BT (r+1)  \, \delta  U^\beta \right]=0  
\end{split}
\end{align}
and the second one of these equations contracted by $h_\beta^\delta$ and $U_\beta$ give respectively
\begin{align}\label{wave3}
\delta  U^\delta = \frac{-c \,  \delta  (nT)}{\lambda n T (r+1)} \, \nu^\delta   \quad , \quad  \delta  (nT) = \frac{n T}{\lambda c} \, \frac{r+1}{r} \, \nu_\alpha  \, \delta   U^\alpha \, .
\end{align}
By substituting \eqref{wave3}$_1$ in \eqref{wave2}$_1$ and \eqref{wave3}$_2$, we obtain 
\begin{align*}
\delta  n = \frac{1}{\lambda^2 T (r+1)} \, \delta  (nT) \quad , \quad  \left(\lambda^2 - \frac{1}{r} \right) \, \delta  (nT) =0 \, .
\end{align*}
Hence, if $\lambda^2 \neq {1}/{r}$, we find $\delta n =0$, $\delta T =0$,
$\delta  U^\delta =0$. \\
If $\lambda^2 = {1}/{r}$, we have the following unique condition on $\delta n$ and $\delta T$
\begin{align*}\label{wave4}
\frac{\delta  n}{n} = r \, \frac{\delta  T}{T} \quad \mbox{ and, moreover,}  \quad \delta  U^\delta = \frac{-c \nu^\delta   }{\lambda \, n \, r} \,  \delta  n .
\end{align*}
We conclude  that $\lambda  = \pm \sqrt{{1}/{r}} = \pm \sqrt{p/e}$  are two  eigenvalues with multiplicity one. Jointly with  $\lambda=0$ we have a set of five independent eigenvectors and this fact proves the hyperbolicity of the system in the ultra-relativistic limit. Therefore taking into account the  Statement 1  the Statement 2 is proved. We observe that the characteristic velocities in light speed unity are the same of the one of monatomic gas until $-1 \leq a\leq 2$ and for $a>2$ the non null velocities decay as a square root of $1/(a+1)$ when $a$ increase.
	
The ultra-relativistic limit for the causal dissipative  full system proposed in \cite{Pennisi_Ruggeri} will be published elsewhere soon.

\end{document}